# Functional Parcellation of fMRI data using multistage k-means clustering

**Harshit Parmar,**[a*] **Brian Nutter,**[a] **Rodney Long,**[b] **Sameer Antani,**[b] **Sunanda Mitra**[a]

[a] Department. Of Electrical and Computer Engineering, Texas Tech University, Lubbock Texas, USA
[b] Lister Hill National Center for Biomedical Communications, National Library of Medicine, National Institutes of Health, Bethesda, MD, USA

**Abstract:**

Purpose: Functional Magnetic Resonance Imaging (fMRI) data acquired through resting-state studies have been used to obtain information about the spontaneous activations inside the brain. One of the approaches for analysis and interpretation of resting-state fMRI data require spatially and functionally homogenous parcellation of the whole brain based on underlying temporal fluctuations. Clustering is often used to generate functional parcellation. However, major clustering algorithms, when used for fMRI data, have their limitations. Among commonly used parcellation schemes, a tradeoff exists between intra-cluster functional similarity and alignment with anatomical regions.

Approach: In this work, we present a clustering algorithm for resting state and task fMRI data which is developed to obtain brain parcellations that show high structural and functional homogeneity. The clustering is performed by multistage binary k-means clustering algorithm designed specifically for the 4D fMRI data. The results from this multistage k-means algorithm show that by modifying and combining different algorithms, we can take advantage of the strengths of different techniques while overcoming their limitations.

Results: The clustering output for resting state fMRI data using the multistage k-means approach is shown to be better than simple k-means or functional atlas in terms of spatial and functional homogeneity. The clusters also correspond to commonly identifiable brain networks. For task fMRI, the clustering output can identify primary and secondary activation regions and provide information about the varying hemodynamic response across different brain regions.

Conclusion: The multistage k-means approach can provide functional parcellations of the brain using resting state fMRI data. The method is model-free and is data driven which can be applied to both resting state and task fMRI.

*Harshit Parmar, email: harshit.parmar@ttu.edu

## 1 Introduction

Functional Magnetic Resonance Imaging (fMRI) is a non-invasive technique used to study the neural activation dynamics of the brain. A localized neuronal activation causes an increase in blood flow and blood oxygenation concentration in that region. The difference in the blood oxygenation causes a change in the local magnetization that can be detected in the MR s ignal [Bandettini et al., 1994; Ogawa et al., 1990]. Since fMRI measures blood oxygenation (BOLD: Blood Oxygenation Level Dependent) changes following a neuronal activation, it is an indirect measure of neuronal activity. fMRI studies can be broadly classified into two main types, task fMRI and resting-state fMRI. In task fMRI, the main aim is to identify regions of the brain corresponding to different physical and mental activities [Bandettini et al., 1992]. In rest fMRI, the main aim is to study the underlying spontaneous activation and communication between different brain regions [Biswal et al., 1995]. One of the methods to determine the communication among regions is by using the temporal correlation between the fMRI time series of different brain regions [Van Den Heuvel & Hulshoff, 2010]. Because of the large number of voxels in the brain acquired by fMRI, voxel-wise correlation calculation is highly computationally expensive and not practically feasible. To overcome the limitation of voxel wise correlation computation, earlier analyses used a seed-based approach, [Biswal et al., 1995] where a random seed-voxel is selected within the brain, and a region is formed by combining all the voxels that have a high temporal correlation with the seed voxel time series. Because the seed selection is arbitrary, the regions obtained in such a fashion are highly dependent on the initialization of the seed voxel. As an alternative, researchers perform clustering to obtain different brain regions. A region-wise analysis is then performed with predefined regions. One main challenge is to obtain homogenous brain parcellation from the resting state data.

Several brain parcellation schemes have been discussed in the literature [Thirion et al., 2014], but each has its advantages and disadvantages. When resting-state fMRI was being developed, anatomical atlases were commonly used. It was believed that regions having anatomical similarity might also have functional similarity [Nieto-Castanon et al., 2003]. In this technique, the time series of all the voxels within a given anatomical region are averaged to obtain a representative time series for that region. In reality, some anatomical regions are small and may be functionally similar, while certain anatomically large regions, such as Brodmann regions 6, 11, 37 and 48, may not be functionally homogenous and have subdivided or overlapping functional regions. For those large anatomical regions, such a representative time series is not always an appropriate descriptor since the region itself is not functionally homogenous, and temporal variations within different subregions are lost due to averaging, thereby yielding inaccurate analysis. Some researchers also use data-driven techniques like Independent Component Analysis (ICA) [Beckmann et al., 2005; Calhoun et al., 2001a; Calhoun et al., 2001b], mixture models [Golland et al., 2007], graph-based approach [Shen et al., 2010], dictionary learning [Wang et al., 2016], etc. to obtain functionally homogenous parcellation of the brain. Sometimes, clusters obtained by such an approach may lack spatial homogeneity as the clusters may be scattered across the brain regions. To have spatially uniform parcels, functional atlases based on spatially constrained spectral clustering [Craddock et al.,2012], and seed-based region growing [Blumensath et al., 2013], have been developed. The main limitation of a global atlas is the inability to account for subject-level differences. Moreover, it is important to note that atlas-based approaches may have alignment and boundary issues with improper coregistration scheme.

To address these shortcomings, we propose a novel multistage k-means clustering algorithm that can be used to generate spatially and functionally homogenous brain regions from fMRI data. The multistage k-means approach is an unsupervised clustering approach. Because no mapping of regions from a separate atlas is required, the proposed approach avoids the problem caused by inappropriate coregistration. Moreover, the algorithm takes advantage of both spatial and functional constraints to obtain spatially and functionally homogenous parcels. The two of the most commonly used clustering algorithms are k-means clustering [Hartigan & Wong, 1979] and hierarchical clustering [Johnson, 1967]. Both k-means and hierarchical are unsupervised clustering algorithms, but when applied to fMRI data, they have their limitations. In k-means, we start by selecting random centroids within the dataset. Each data point is assigned to the cluster with the shortest distance between the data point and cluster centroid. After all the data points have been assigned to a cluster, new corresponding cluster centroids are computed, and the process is repeated until the change in cluster centroid position for all clusters is less than some predefined threshold. For fMRI data, the clustering results can easily become biased due to arbitrary cluster centroid initialization. In hierarchical clustering, we start by computing the pairwise distance matrix between every pair of data points. Initially, each data point is treated as a separate cluster. Based on the pairwise distances, the closest data point pairs are merged until all data points finally belong to a single large cluster. The entire merging process is represented by a hierarchy tree known as the dendrogram. Different clusters are obtained by splitting the dendrogram at a different distance threshold. Because of the large number of voxels in the fMRI data, typically ranging from 50,000 to 200,000 voxels per brain volume, the calculation of the pairwise distance matrix for hierarchical clustering is computationally impractical.

The proposed clustering approach combines the advantage of both hierarchical and k-means clustering approach for fMRI dataset while reducing the limitations. Our method supports the notion of a hierarchical structure. Within each stage or level of the hierarchy, the clustering is done based on the k-means algorithm. The level-wise clustering avoids the computation of a full distance matrix, and the hierarchical structure gives the flexibility to get any desired number of clusters. Each k-means clustering step is performed to obtain at most 2 clusters. A small number of clusters reduces the bias of centroid initialization and yield stable clusters. The multistage k-means approach is also not limited to resting-state fMRI. It can be used with task fMRI data as well to get information about activation regions and the shape of the Hemodynamic Response Function (HRF). The advantage of a model-free approach in estimating the shape of the hemodynamic response eliminates the bias introduced because of an HRF model. Moreover, different clusters in task fMRI data can be used to investigate more complicated issues like the variability of the hemodynamic response [Handwerker et al., 2004; Handwerker et al., 2012] and the whole-brain activation [Gonzalez-Castillo et al., 2012].

**2. Materials and Methods**

*2.1 Multistage k-means Algorithm*

The multistage k-means clustering algorithm is used for functional brain parcellation using fMRI data. The clustering is performed on the preprocessed 4D fMRI time series data. Before applying the clustering algorithm to fMRI data, the preprocessed 4D fMRI data structure is modified. The preprocessed data is vectorized into a 2D matrix whose rows correspond to voxels and whose columns correspond to time points. The original 4D fMRI volume is converted to an N x T matrix, where N is the total number of voxels inside the brain region, and T is the number of time points. Thus, each of the N voxel's time series is treated as a T dimensional feature vector.

All feature vectors start as a single parent cluster. Before clustering is begun, the desired number of hierarchical stages and a correlation threshold is specified, denoted by 'NS' and 'CT', respectively. The parent cluster is then split into two children clusters using k-means. The k-means split also yields cluster centroids. For fMRI data, the centroids are T-dimensional and are considered as representative time series for the clusters. Next, a correlation is computed between the representative time series of the children clusters. If the correlation between the representative time series of the children clusters is greater than the threshold specified, it is assumed that the child clusters are a part of the same cluster and further clustering of the parent cluster is stopped. The parent cluster is considered as a converged cluster and saved for further use. Such correlation stopping condition prevents the subdivision of a single cluster into multiple sub-clusters. However, if the correlation between children clusters is less than the threshold, then each of the children clusters is considered as an independent parent cluster in the next level of the hierarchy. This process is repeated until all the parent clusters have converged or the last stage of the hierarchy is reached. Figure 1 shows the pseudocode for the multistage k-means algorithm. A dummy example for multistage k-means is shown in Appendix A1.

In k-means clustering, the most common distance metric is the Euclidean distance. In this approach, however, a correlation-based distance metric has been used. Because for functional parcellation, the aim is to combine voxels having similar temporal fluctuation, a correlation distance metric is preferred over Euclidean. Moreover, the maximum number of iterations for each k-means split is set to 100 with 5 replicates. The iteration corresponds to a single centroid update in the k-means algorithm while replicates correspond to repeating the entire clustering processing. The centroid initialization is done using the k-means++ algorithm of MATLAB. At the end of the last stage of the hierarchy, pairwise temporal correlation is computed between centroids of all the converged clusters. Any cluster pair for which the correlation coefficient is higher than the correlation threshold specified earlier is merged.

*2.2 Similarity Index Measure*

A measurable metric to test the similarity between two different spatial clusters was used by Mezer et al. [Mezer et al., 2009]. This metric is called the similarity index analysis. To compare the similarity between spatial clusters obtained by the multistage k-means approach to existing functional and anatomical atlases, we used a slightly modified form of the similarity index analysis. The mathematical form to compute the cluster wise similarity index ($SI_i$) and the maximum similarity index (mSI) between two different cluster sets α, and β is given in equations (1) and (2):

$$SI_i = \max_j \sqrt{\frac{|\alpha_i \cap \beta_j|^2}{|\alpha_i| \cdot |\beta_j|}} \quad (1)$$

$$mSI = \max_i (SI_i) \quad (2)$$

Here, i = 1, 2, …, Nα and j = 1, 2, …, Nβ, with Nα and Nβ are the maximum number of clusters in α and β, respectively. αi and βj are the number of elements in the i$^{th}$ and j$^{th}$ clusters of α and β, respectively. αi ∩ βj is the number of elements in αi that overlap with elements of βj. For two identical datasets, the mSI value is 1. Moreover, for two datasets with a different number of clusters, the mSI value is always less than 1. Thus, if Nα ≠ Nβ, mSI < 1.

Because the ordering of the cluster number may be different in different sets, the first step is to find the closest cluster in set β corresponding to each cluster in set α. For each cluster in α, similarity index is computed with every other cluster in β. The cluster in β having the highest similarity index measure for cluster αi is considered the closest cluster, and the similarity index measure is assigned to that particular cluster. This process is represented by equation (1). For every cluster in α, the closest cluster is detected, and the similarity index measure is computed. To compute the similarity index between different atlases and spatial maps, it is necessary that the physical scaling of both the spatial maps be the same. To make sure that the spatial maps being compared are of the same size, a 3D affine transformation is applied using the 'Coregister (Estimate and Reslice)' option of the Statistical Parametric Mapping (SPM) 12 MATLAB toolbox [https://www.fil.ion.ucl.ac.uk/spm/software/spm12/]. The Brodmann atlas [Maldjian et al., 2003] and Craddock atlas [Craddock et al.,2012] were transformed to match the size of spatial maps obtained by clustering. The method of interpolation was chosen to be nearest neighbor to avoid the addition of interpolated intensity values and have sharp and definite boundaries between the regions.

---

**Multistage k-means Algorithm pseudo code:**

| | | |
|---|---|---|
| **Data:** | $f_{mat}$ | // preprocessed fMRI data (vectorized) |
| | $CT$ | // correlation threshold |
| | $NS$ | // number of stages in hierarchy tree |
| **Output:** | $K$ | // clusters |
| | $t_{mat}$ | // representative time series |

```
/* Beginning multistage algorithm */
/* Initialize parameters */
n = 1                                    // set current hierarchy level to 1
PCⁿ ← f_mat                              // add f_mat to list of parent clusters (PC)

/* Check for last stage of hierarchy or convergence of all clusters */
if ((n ≤ NS) and (not_empty (PCⁿ))):
    for i in PCⁿ:
        [A, B] ← kmeans (PCⁿᵢ, 2)        // split parent cluster into two subclusters
        if (correlation (A, B) ≥ CT):    // check for similarity between subclusters
            
            /* If similarity between subcluster greater than threshold */
            K ← PCⁱ                       // Parent cluster converged, so save it
        else:
            
            /* If similarity between subclusters less than threshold */
            PCⁿ⁺¹ ← [A, B]                // Each subcluster is parent cluster for next stage
        end
    end
    n += 1
end

/* Merge similar clusters in K */
for i in K:
    for j ≠ i:
        if (correlation (Kᵢ, Kⱼ) ≥ CT):
            K ← merge (Kᵢ, Kⱼ)            // Merge similar clusters
        end
    end
    t_mat ← centroid (Kᵢ)                 // Save representative time series for cluster
end
```

---

*2.3 Dataset and Preprocessing*

The performance of the multistage k-means was first validated on a synthetic time series data. The synthetic ground truth data consists of 8 x 8 2D grid with six distinct regions. For each region, a representative time series was obtained by sampling a sinusoid with 50-time points. The sampling frequency was 10 Hz, and the frequencies for each region were 0.65 Hz, 0.7 Hz, 0.7 Hz, 0.8 Hz, 0.5 Hz, and 0.9 Hz with a random phase. Entire synthetic time series data was created by assigning the corresponding representative time series to each 8 x 8 pixel and then adding Gaussian random noise with different variance. Figure 2A shows the 2-D ground truth regions a sample visualization of the synthetic time series data. Four different synthetic datasets were created using different noise variance to obtain a Signal to Noise Ratio (SNR) value of 2, 1, 0.5, and 0.25. The structure of the dataset was such to have spatiotemporal data similar to 4D fMRI.

Next, a simulated resting-state fMRI data was generated using f-Sim fMRI simulator [Parmar et al., 2018]. The simulated resting-state fMRI data was generated using a multivariate covariance matching scheme with Craddock atlas as the functional reference parcellation. The spatial and temporal properties of the simulated data were similar to an ordinary resting-state fMRI scan. The data were simulated for a 240 s long resting-state experiment with a Repetition Time (TR) of 2 s yielding 120 brain volumes. Each brain volume consists of 48 axial slices of thickness 3.5 mm with an in-plane matrix size of 64 x 64 and a resolution of 3 mm x 3 mm.

The simulated data validation acts as a bridge between the synthetic and the real data. The simulated data has the structure of real fMRI data with an added advantage of having the ground truth to compare with. For the simulated data, the correlation threshold was set to 0.7 as the stopping criteria with a maximum of 7 stages in the hierarchy. The selection of correlation threshold was arbitrary based on the results from the synthetic data output. The number of stages was set to 7, which may yield a maximum of 128 clusters. The simulated data was generated using the Craddock atlas with 50 parcellations. Thus, the number of stages were set to a power of 2, which can produce more than 50 clusters but not the immediate next. Hence NS was set to 7 and not 6.

The resting-state fMRI dataset was obtained from the 1000 Functional Connectome Project (FCP) dataset [http://fcon_1000.projects.nitrc.org/index.html] by the Child Mind Institute and the International Neuroimaging Data-Sharing Initiative (IDNI). The results shown here are for the Pittsburgh dataset, which consists of 17 subjects [10M/ 7F; age 25 – 54 years]. One of the subject's data was excluded from analysis due to the presence of excessive motion. The functional and anatomical scan was taken on a 3T scanner. The total scan time for functional data was a little less than 7 minutes and with a TR of 1.5 s yields a total of 275 time points. Each 3D volume consists of 29 axial slices with a slice thickness of 3.2 mm. The in-plane matrix size is 64 x 64, with a resolution of 3.125 mm x 3.125 mm. The structural scan consists of 224 sagittal slices of thickness 0.78 mm. The in-plane matrix size is 256 x 256, with a resolution of 0.78 mm x 0.78 mm. Additional details can be obtained on the 1000 FCP website [http://fcon_1000.projects.nitrc.org/fcpClassic/FcpTable.html]. For resting state data for all 16 subjects, the performance of the multistage k-means clustering output was compared with simple k-means, ICA clustering and Craddock functional atlas using student's t-test.

The task fMRI dataset was obtained from the 100 functional runs per subject dataset collected at the National Institute of Health (NIH) by Dr Javier Gonzalez-Castillo [https://central.xnat.org/app/action/DisplayItemAction/search_value/100RunsPerSubj/search_element/xnat:projectData/search_field/xnat:projectData.ID]. The dataset was used to study the whole brain time-locked activation by massive averaging. The functional experimental paradigm was a block design with five blocks with a visual stimulus of 20 s ON and 40 s OFF. It also had 30 s rest period before and 10 s after the main experiment paradigm. The TR was 2 s, resulting in 165 time points in total for each time series. Each 3D volume consists of 45 axial slices of thickness 3.75 mm. The in-plane matrix size is 68 x 71 with a resolution of 3.75mm x 3.75 mm. The structural scan consists of 124 axial slices with a slice thickness of 1.2 mm. The in-plane matrix size is 256 x 256, with a resolution of 0.9375 mm x 0.9375 mm. Additional details on the dataset can be found in Gonzalez-Castillo et al., 2012.

A standard preprocessing pipeline was used to preprocess the fMRI dataset. All the preprocessing was done using SPM12 toolbox [https://www.fil.ion.ucl.ac.uk/spm/software/spm12/] [Ashburner et al., 2016] and MATLAB. The preprocessing steps include motion correction, coregistration, normalization, smoothing, temporal signal drift and global signal regression, and brain segmentation. For motion correction, all the functional volumes were realigned to the mean volume and 6 motion parameters, 3 translation and 3 rotational, were corrected for. The interpolation was done using the 4th order B-spline method. The functional and anatomical volumes were coregistered with one another and were then normalized to the standard MNI space. The normalized functional volumes were smoothed using a 3D Gaussian kernel with a Full Width Half Maxima (FWHM) of 8 mm. Temporal signal drift and global signal fluctuations were estimated and reduced using a PCA drift removal algorithm [Parmar et al., 2019]. The PCA algorithm uses eigenvalue decomposition of the entire time series data to estimate the low frequency drift and global signal fluctuations. The eigenvalue reconstruction of these noise components is then subtracted from the time series of all voxels to reduce the effect of temporal signal drift and global signal fluctuation. Without global signal regression, the clustering output depends on the absolute intensity (global signal) and not the temporal fluctuation. The BOLD fluctuations are less than 2%-5% the absolute signal intensity and thus global signal regression is an essential preprocessing step. Lastly, the voxels outside the brain region were thresholded and masked out. Masking reduces the effective number of voxels and hence the complexity. Moreover, for both task and resting-state fMRI data results shown in the paper, the multistage clustering was applied with CT = 0.7 and NS = 7 unless specified otherwise. The value for CT and NS were selected based on the results of simulated fMRI data.

**3. Results**

*3.1 Synthetic Data*

First, the multistage k-means clustering algorithm was validated using the synthetic data. Clustering output was computed for each of the four cases, different SNR values. The clustering output was compared with the ground truth using the mean similarity index measure. For the multistage k-means, the correlation threshold (CT) stopping criteria was set to 0.7. The ground truth matrix has 6 clusters, so the number of stages (NS) was set to 3, the next highest power of 2 closest to 6. The clustering output of the multistage k-means was also compared with the clustering output of simple k-means. For comparison to be uniform, the same number of clusters were used in simple and multistage k-means. To have a uniformity in the cluster numbers, the multistage approach was run first, and the number of clusters were determined. Then simple k-means was run to get the same number of clusters in the output. Figure 2C shows the clustering output for both the approach for different values of SNR and different combination of CT and NS. Different colors in the 2D matrix represent different clusters. The accuracy of the clustering was computed using the mean of the similarity index measure for all clusters. As expected, the clustering accuracy reduces with reducing SNR value. Also note, the mean similarity index for multistage k-means is always slightly better than simple k-means. Along with that, the number of clusters detected by the multistage approach increases with reducing SNR. With increasing noise, the intra-cluster correlation reduces and hence some of the clusters that converge for low noise variance fail to converge for a higher noise variance.

Next, the effect of NS was investigated by repeating the clustering with 4 stages in the hierarchy instead of 3. The clustering results for that is shown in Figure 2D. For lower noise variance, SNR = 2, as all the clusters converge before reaching the maximum number of stages, the clustering output remains the same. However, for larger noise variances, the clusters do not converge until the last stage of the hierarchy and thus an increase in the total number of clusters in observed. The increase in the total number of clusters also reduces the similarity index as compared to lower NS. Nevertheless, for this case also, the multistage approach performs slightly better than the simple k-means.

Lastly, the effect of CT was computed using synthetic data. Clustering was performed for a higher, 0.85, and a lower, 0.4, value of CT keeping the NS as 3 is shown in Figure 2E and 2F. For a lower value of CT, there is a higher probability that the clusters will converge because of the low threshold. Higher convergence probability results in a lower number of total clusters obtained. It can be noted from the results that even for a lower SNR, 0.25, clusters converge, and the total number of clusters obtained is less than the maximum number of cluster possible. One more thing to note is that for a lower CT, simple k-means performs better than the multistage counterpart. On the other hand, a higher value of CT means fewer clusters will converge because of a tighter threshold. Thus, even for high SNR, 2, clusters will fail to converge and will result in a higher number of clusters.

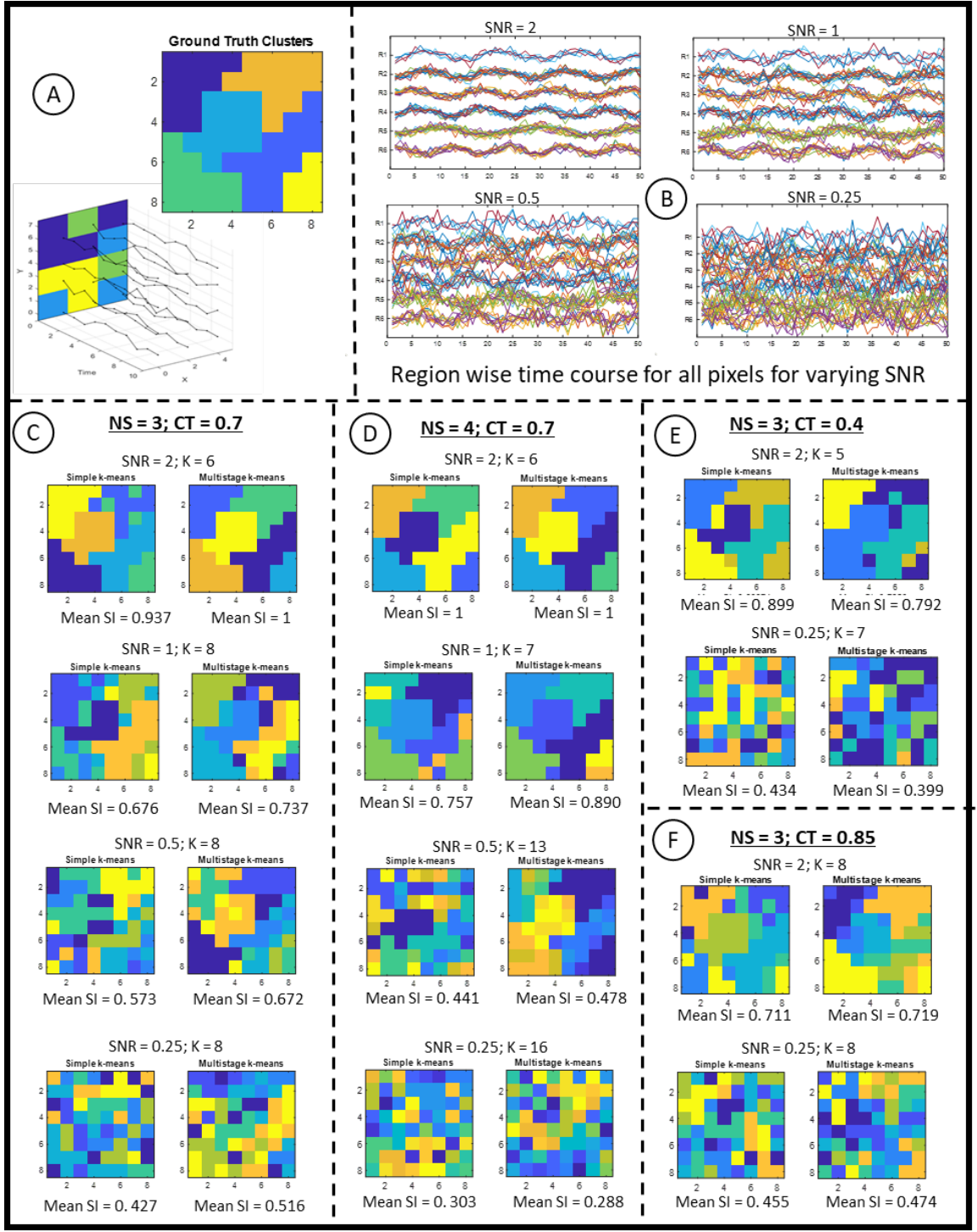

Figure 2: (A) 2D ground truth matrix and visualization of synthetic 2D time series data. (B) Noisy time series with different SNR for voxels in different regions. (C), (D), (E) and (F) Results for synthetic time series data. Comparison of clustering output for multistage (right) and simple k-means (left) for different values of SNR, CT and NS. Different colors in the 2D matrix represents different clusters. The total number of clusters identified by the multistage algorithm is represented by 'K' on top of the plots. Mean similarity index values shown below each visualization.

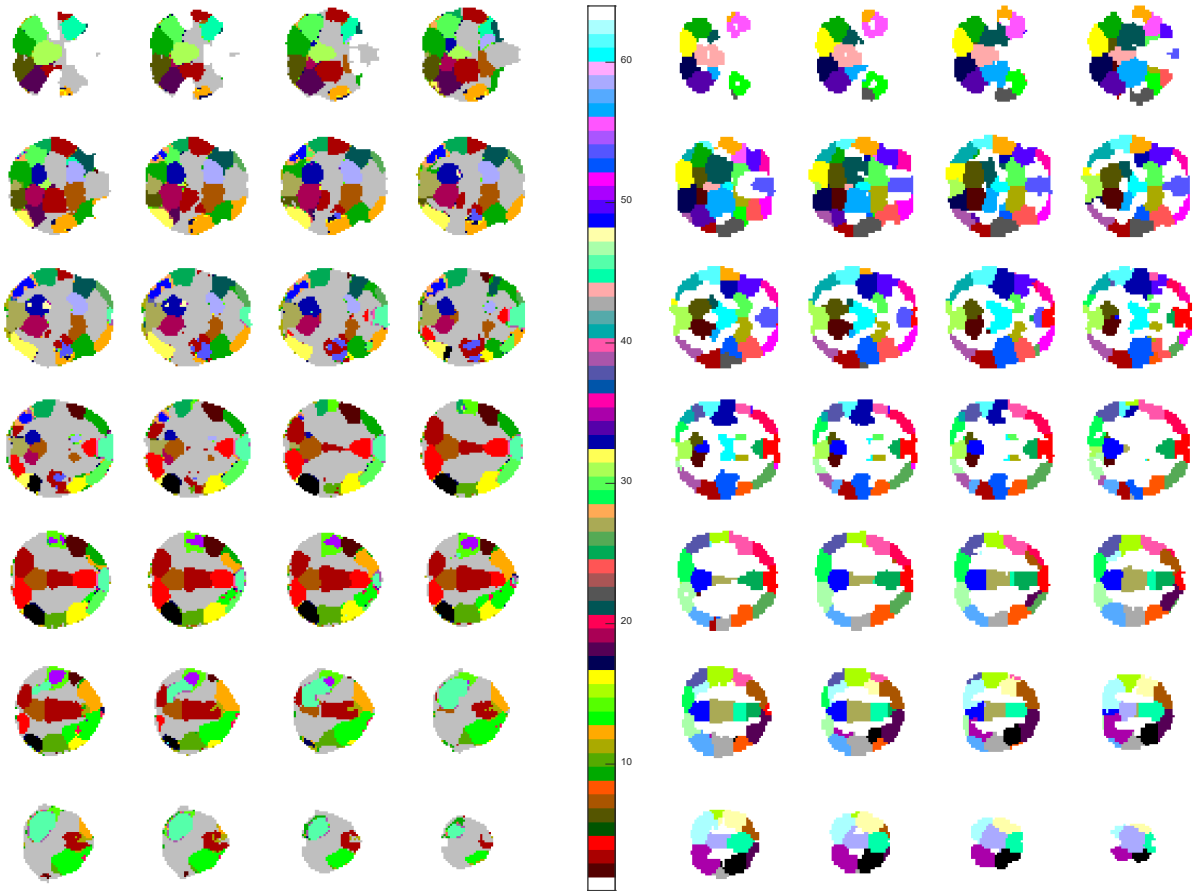

Figure 3: (Left) Spatial clusters for multistage k-means. (Right) Craddock atlas wit 50 parcels. Different colors represent different clusters.

## 3.2 Simulated fMRI Data

Next, the performance of multistage clustering was validated using simulated resting-state fMRI data. The total number of clusters detected by the multistage approach was 54. Thus, the number of clusters obtained was closer to the actual ground truth (50) than the maximum number of clusters possible (128). The number of clusters being closer to 50 rather than 128 indicates that the majority of the clusters converge before reaching the final stage of the hierarchy suggesting stable clustering. Figure 3 shows the spatial cluster comparison for the output of multistage clustering and the Craddock atlas with 50 parcels. Each cluster is represented by a different color. The color-bar represents the corresponding cluster number. The maximum similarity index measure, mSI, is also computed between the clustering output and the Craddock atlas. A mSI value of 0.771 is obtained that indicates high spatial similarity, which is also observed in the figure. The results obtained on synthetic and simulated time series data show the potential of the multistage k-means algorithm for functional parcellation.

*3.3 Resting-State fMRI Data*

The lack of ground truth for resting-state fMRI data makes a quantitative analysis challenging. Here, the performance of multistage clustering is compared with the output of data driven techniques like ICA analysis and simple k-mean clustering. The performance is also compared against global anatomical and functional atlas such as the Craddock functional atlas [Craddock et al.,2012] and the AAL anatomical atlas [Tzourio-Mazoyer et al., 2002; Rolls et al., 2015; Rolls et al., 2020]. The ICA analysis was performed on MATLAB 2020a using the GIFT ICA toolbox [Rachakonda et al., 2007]. The ICA algorithm was set up to obtain 50 independent components. Functional ROIs and representative time series for each ROI were obtained for each of the Independent Components (ICs). In the case of ICA analysis, the time series of each voxel is assumed to be obtained by a weighted combination of different ICs. Thus, a given voxel can belong to mode than one functional ROI. For simplicity, to have a comparable clustering output, each voxel was assigned to the IC with the maximum contribution to that voxel, hard clustering. All the spatial analysis of the ICA analysis were performed using the hard clustered ICA spatial maps. The spatial comparison was performed using the maximum similarity index (mSI) measure. The mSI values, mean ± standard deviation across all 16 subjects, between different data driven and atlases is shown in Figure 4. Along with the mSI values, Figure 4 also shows the t-value for all the cases. A student's t test was performed to test if there was a significant difference in the mSI values when using the multistage k-means clustering algorithm as opposed to the other. Cases for which the mSI value for multistage k-means is significantly higher (p < 0.05) than the other atlas is highlighted in **blue**. The comparison with both functional and anatomical atlas gives an insight into the structural homogeneity of the parcels. If the clusters have very low similarity with both atlases, then it means the clusters are scattered without any uniform spatial structure. If the similarity of the clusters with anatomical atlas is much larger as compared to a functional atlas, then it might indicate that the parcellation is performed based on anatomical differences and not functional differences. A good functional parcellation scheme should have high similarity with both functional and anatomical atlas with a slightly higher similarity with the functional atlas.

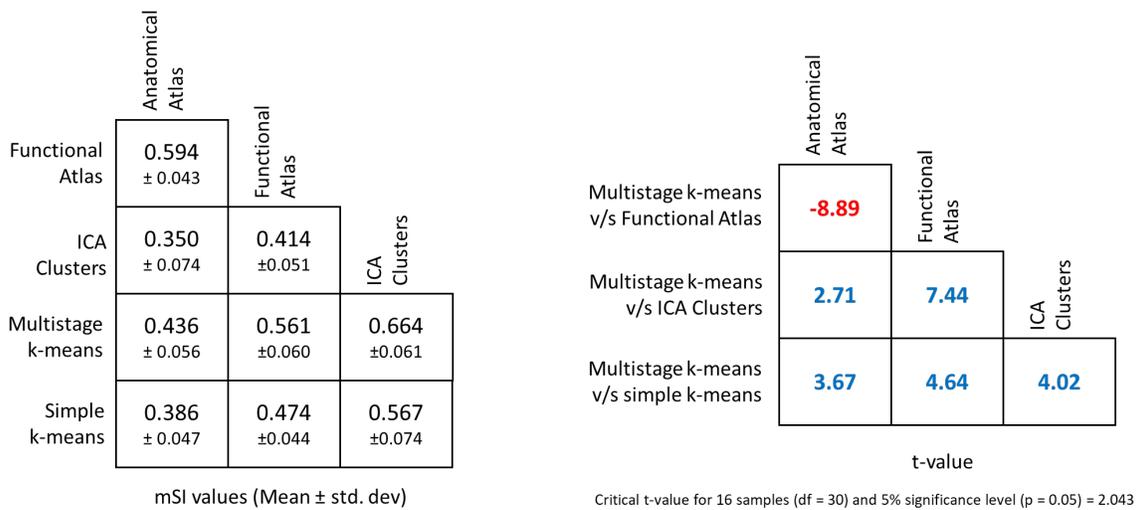

Figure 4: (left) structural similarity measure (mSI) between cluster outputs of different approach averaged (mean ± standard deviation) over 16 subjects. (right) t-value for comparing structural similarity for multistage k-means and different approach. The values in blue indicate significant higher values for multistage while in red indicate significant lower values for multistage approach as compared to others.

At first glance the similarity index values might appear very low but different number of clusters in both the clustering output being compared reduces the mSI values as discussed earlier. Moreover, for reference, the maximum similarity index was computed between two synthetic clusters of size 16 x 16 x 16 with 8 randomly distributed spatial clusters each. The mean mSI value obtained for 100 random simulation was 0.025. Compared to the random simulation, the mSI values obtained for clustering outputs is much higher. The number of output clusters for both k-means approach was 58, the number of ICs were 50 and thus, to have a better comparison, Craddock atlas with 60 parcels was used. The limitation with the anatomical atlas is the fixed number of parcels (100 ROIs inside the brain region); thus, a slightly lower mSI is expected for comparison with anatomical atlas as there exists a difference in the total number of clusters. The multistage approach performs better than simple k-means and ICA in terms of mSI values with both anatomical and functional atlases. Even when comparing with the data driven clusters (ICA output), the multistage approach performs better than simple k-means. Craddock atlas' similarity is higher than the similarity of the multistage clustering output for anatomical atlas. The lowest mSI value is obtained between ICA clusters and anatomical atlas. As discussed earlier, data driven techniques like ICA yields cluster that have high functional similarity compared to structural similarity. The t-test result also suggests that using multistage approach, a significant improvement is observed in structural similarity of the output clusters as compared to other clustering approach and data driven techniques.

The spatial clustering output for simple and multistage k-means and ICA is shown in Figure 5. Each cluster is represented by a different color. The color-bar indicated the color corresponding to different cluster number. Visually, the clusters for both the sets of k-means look spatially uniform while the ICA clusters look little scattered. One thing to observe for multistage k-means clusters is a visually apparent difference between the gray matter and the white matter which is less visible in the simple k-means clusters or ICA.

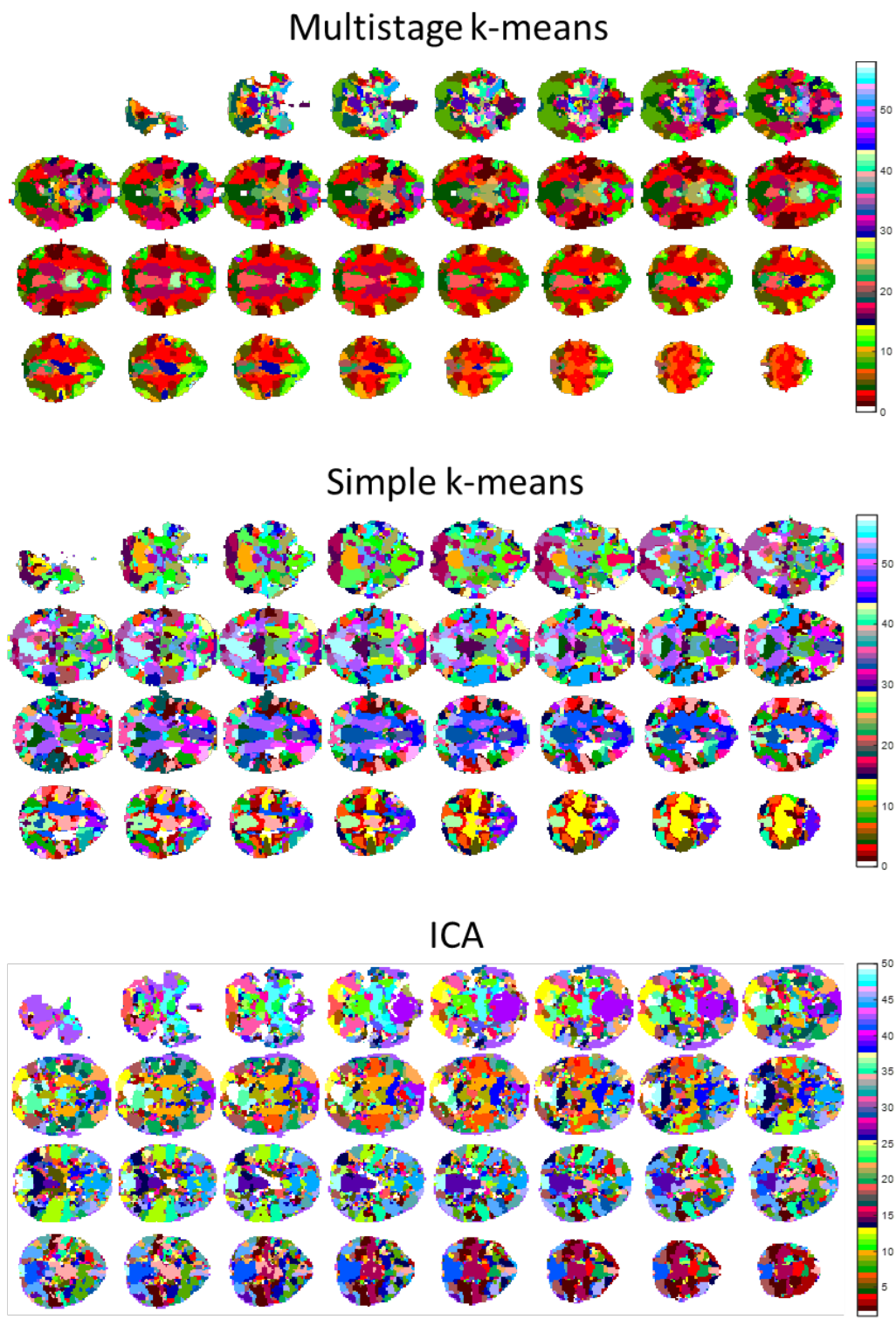

Figure 5: Spatial clustering comparison of resting-state fMRI data for (Top) simple k-means and (middle) multistage k-means and (bottom) ICA. Different colors represent different clusters.

Traditionally, the neuronal activities were only observed inside the gray matter. Recent literatures do suggest white matter activations being observed in fMRI studies. However, the nature of activation is different in white matter and gray matter [Grajauskas et al., 2019]. The gray matter activation is due to post-synaptic potential while that of white matter is due to action potential. A difference in the hemodynamic response of gray matter and white matter is also observed [Li et al., 2019]. Thus, it can be assumed that the temporal response for gray atter and white matter should be slightly different, and a good clustering scheme should be able to identify clusters that belong mostly into either gray matter or white matter.

To check if the clusters belong strictly to gray matter or overlap between gray matter and white matter, percentage overlap was computed for all the clusters to gray matter and white matter segmentation mask. Figure 6 shows the percentage overlap for each cluster to gray matter and white matter in sorted order for multistage & simple k-means and ICA. For multistage k-means, nearly half of the clusters (30 out of 62) have more than 70% overlap with gray matter whereas for simple k-means it is only one third (20 out of 62) and for ICA only one fourth (12 out of 50). Moreover, the clusters that overlap significantly with both gray matter and white matter, i.e. gray matter overlap between 40% and 60%, are almost twice for both simple k-means (15 clusters) and ICA (14 clusters) as compared to multistage (8 clusters).

The overlap results show that for multistage k-means, the parcellations are made based on functional information. For simple k-means, as large number of centroids are initialized at once, more weight is given to spatial similarity. As opposed to that, because parcellations are performed in a hierarchical fashion based on temporal correlation between centroids, the clusters obtained by multistage k-means are both functionally and spatially homogenous. For ICA, however, because a hard clustered spatial map was used, clusters are scattered across the brain as seen in Figure 5.

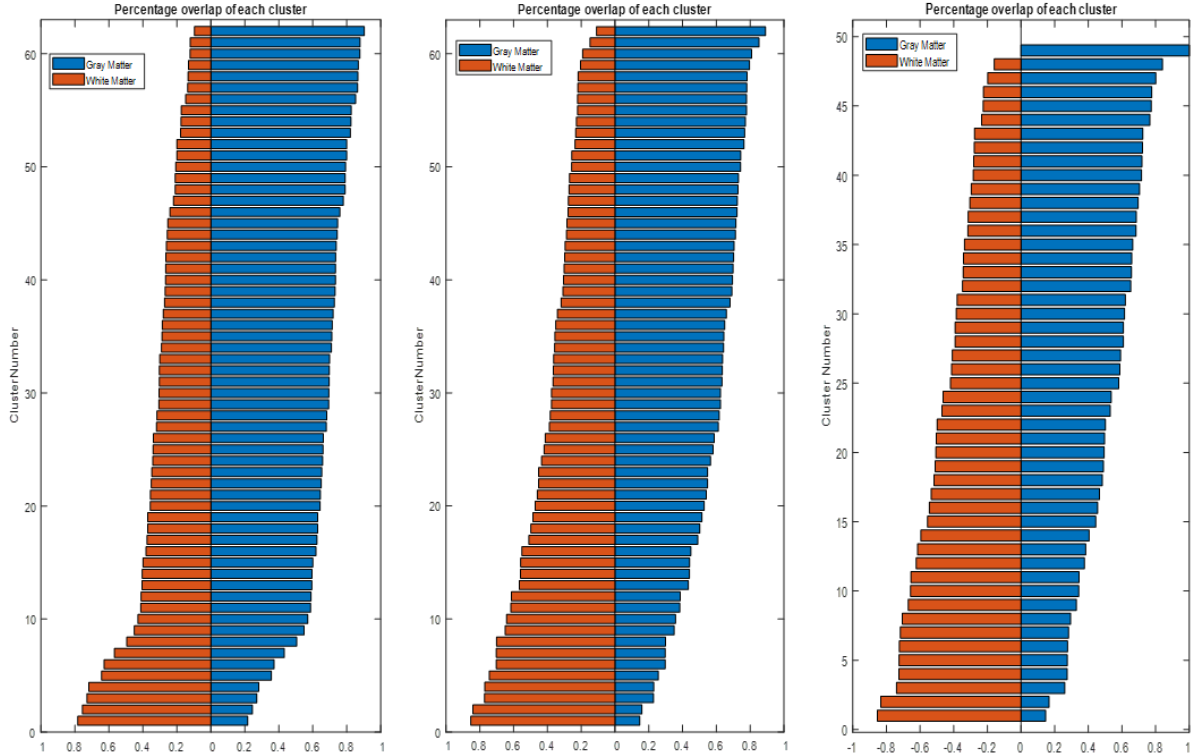

Figure 6: Cluster overlap comparison for output of (left) multistage k-means, (center) simple k-means and (right) ICA. Clusters sorted according to gray matter overlap percentage. The x-axis represents normalized percentage with 0% - 100% scaled to 0 - 1.

The spatial homogeneity can be quantified using the mSI. To quantify the functional similarity, mean intra-cluster correlation histogram was computed. For each cluster, the correlation coefficient is computed between the time series of each voxel with the representative time series. The mean of all the correlation value is computed. A mean correlation value of close to 1 suggests high functional homogeneity within the cluster, while that of close to 0 suggests poor functional homogeneity. In reality, no region of the brain is totally homogenous at the cluster scale in discussion, thus a mean correlation value of 1 is practically not possible. The mean intra-cluster correlation computation is repeated for all the clusters, and a histogram is plotted using all the mean intra-cluster correlation values. Figure 7 shows the mean correlation histogram for multistage k-means, simple k-means, ICA clusters, and Craddock atlas for a single subject.

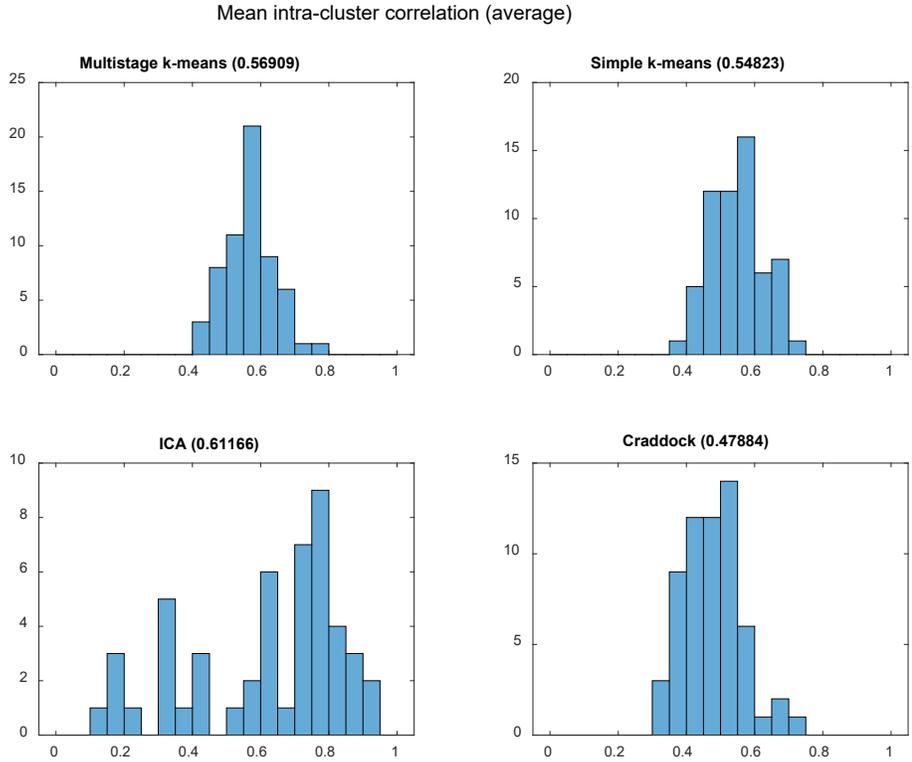

Figure 7: Histogram for mean intra-cluster correlation for different clustering outputs. The average value for the histogram is mention on top of each subplot. The plots show the degree of functional similarity of different spatial clusters.

The maximum functional similarity is observed in the ICA clusters, which is expected. One more observation is the wider range of histogram for ICA as compared to others. Next is the multistage approach which performs better than simple k-means and atlas-based approach. Finally, the simple k-means which shows better functional similarity than the atlas based parcellation. The atlas-based approach does not account for subject-level differences, which might be the reason for lower intra-cluster functional homogeneity. Figure 8 shows the spatial visualization for some of the clusters that correspond to some of the standard resting-state brain networks. The visualization shows the clusters overlaid on the anatomical scan for one of the randomly selected subject. The most commonly identifiable networks from the figure include visual, default mode, primary motor, auditory, frontal, executive, sensorimotor and cerebellar network.

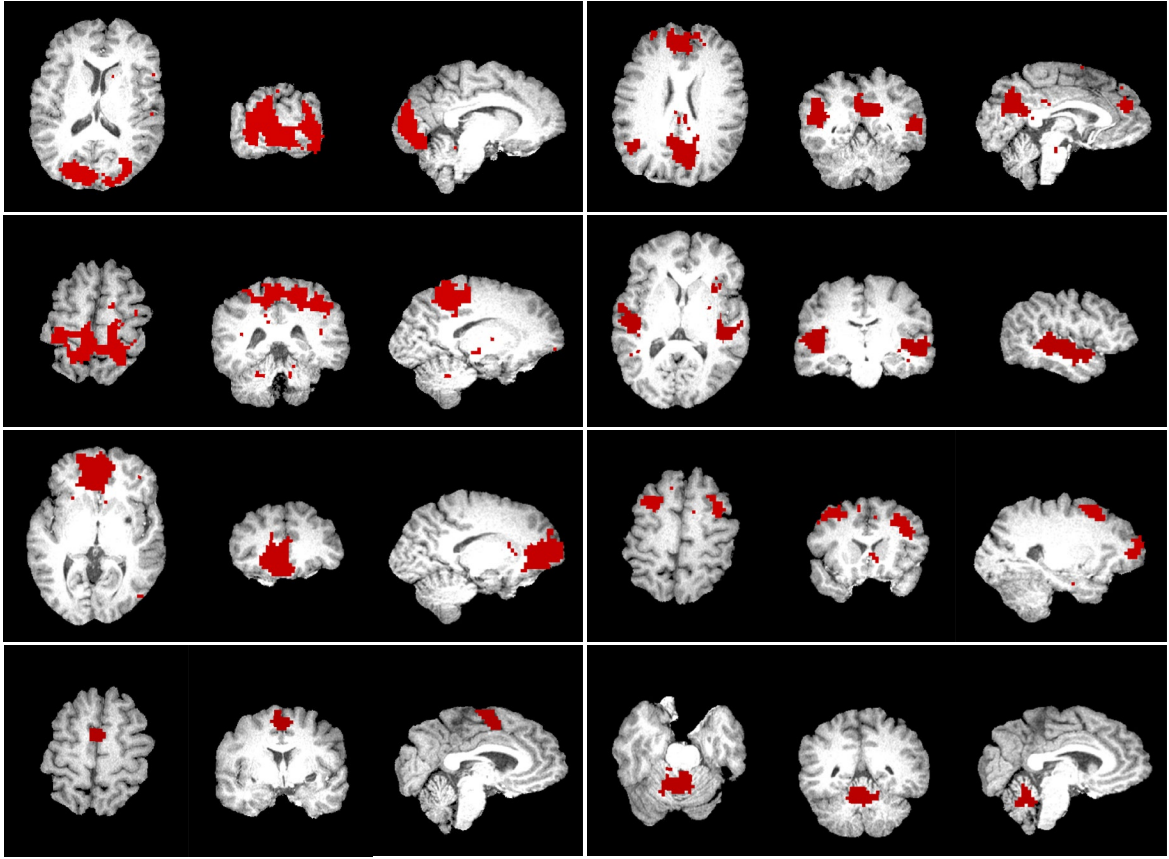

Figure 8: Extracted resting state brain networks using multistage k-means approach. The brain networks (red) are overlaid on the anatomical image (grayscale). The different brain networks shown here are (left to right, top to bottom) visual, default mode, primary motor, auditory, frontal, executive, sensorimotor and cerebellum.

*3.4 Task fMRI Data*

Task fMRI analysis is mainly done for identification of brain regions responsible for different tasks. Multistage k-means clustering can be applied to task fMRI data to obtain the activation region through clustering. For task fMRI, the voxels inside the activation region would have similar temporal fluctuation and hence would be grouped in a single cluster. It has been shown that functional clustering of noise reduced task fMRI data can be used to obtain information about the region having time-locked temporal fluctuation with the experimental paradigm [Gonzalez-Castillo et al., 2012]. Apart from the primary activation region, many brain regions have temporal changes like transients and negative response, which are influenced by the experimental paradigm. Such regions can be defined as secondary activation region. Multistage k-means clustering can be used to identify primary and secondary activation regions and the corresponding hemodynamic response.

The multistage approach identified a total of 32 clusters for the task fMRI data. Figure 9 shows the spatial clusters obtained from the multistage k-means approach. The spatial clusters for task and resting state are different. A possible reason for that is the difference in the temporal BOLD response for task and resting state fMRI. In resting state fMRI, the BOLD response mainly consists of the spontaneous fluctuation inside different brain regions. In task fMRI, the regions that respond to the task have a task specific BOLD response. For example, in a visual task, regions of the visual cortex, like V1, V2 and V3, might have similar temporal response correlated with the task experiment paradigm. Thus, having higher correlation between the time series, all V1, V2 and V3 might be clustered together but during resting state, each region in V1, V2 and V3 might have different temporal response and thus belong to separate spatial clusters. Thus, because the temporal response changes from resting state to task, the correlation of the time series with other regions also changes resulting in different parcellations.

For each cluster in task fMRI, the hemodynamic response was computed using the Finite Impulse Response (FIR) model [Goutte et al., 1999]. For the given data, there were 5 blocks each of 60 s with a TR of 2 s, which yields 30 time points per block. Thus, in the design matrix for the GLM analysis, we have 30 regressors with one corresponding to each time point in the HR. The representative time series of each cluster is used as the target time series to fit the regression model.

The weights corresponding to each regressor correspond to the amplitude of the HR at that time point.

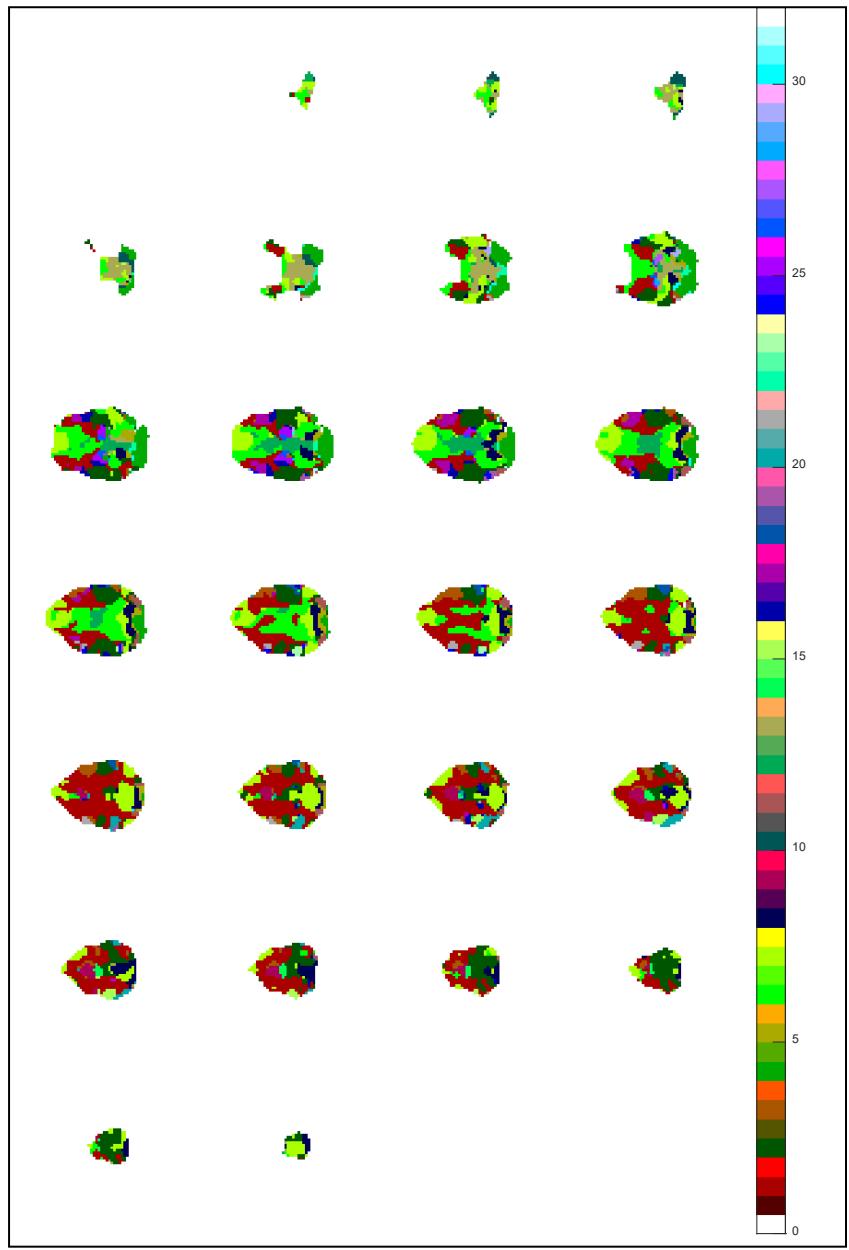

Figure 9: Spatial clusters for task fMRI data obtained using multistage k-means. Different colors represent different clusters.

Figure 10 shows the full hierarchy tree for the multistage approach, with the computed hemodynamic response of each converged cluster in the tree. It can be noted from the hierarchy tree that the clusters 1 to 22 passed the correlation threshold stopping criteria and converged before reaching the last stage of the hierarchy. The representative time series of cluster 4 has the highest temporal correlation with the experimental paradigm. Therefore cluster 4 is identified as the primary activation region. The hemodynamic response of cluster 4 also has the highest absolute signal change. Clusters 3, 5, 9, 11, 16 and 17 also have a positive time-locked response to the experimental paradigm and are considered as secondary activation region. Clusters 24, 27, and 28 also have a positive time-locked response, but because those clusters did not converge until the last stage of hierarchy, those are not considered as secondary activation regions. One more point to note is that after the first split, all the clusters in the right subtree have a negative response to the experimental paradigm while all the clusters in the left subtree have a positive response. Such difference suggest that the partitioning is performed based on larger changes for earlier stages and for later stages the partitioning is done based on finer differences.

Figure 11 shows the spatial visualization of different clusters overlaid on the high-resolution structural scan, along with the corresponding hemodynamic response. The cluster corresponding to the primary activation region, cluster 4, falls within the visual cortex. Given the nature of the experimental paradigm being a visual task, the cluster location is consistent with it. Some of the clusters corresponding to the secondary activation region are also shown. One interesting observation is made for cluster 7. The spatial map for cluster 7 overlaps with the resting state default mode network. The hemodynamic response for this cluster shows anti-correlation with the experiment paradigm, i.e. the absolute intensity decreases during the task period. Previous study has shown that for task fMRI, the default mode network shows anti-correlation with the experiment paradigm [Rombouts et al., 2005], and the result shown here is consistent with that claim. The hemodynamic response of cluster 12 does not have any significant correlation with the experimental paradigm, and spatially the cluster location is outside the gray matter region where neuronal activity is not expected.

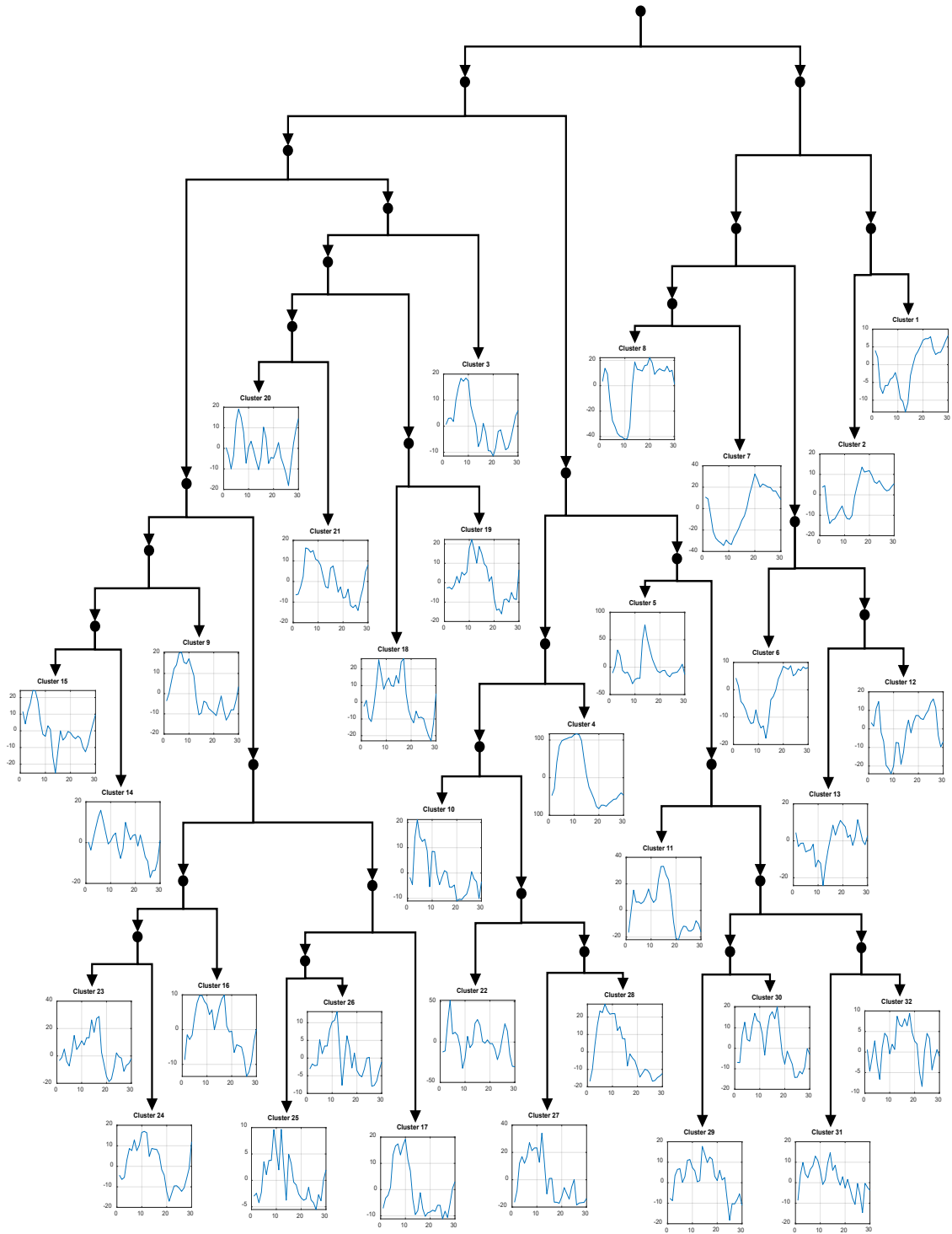

Figure 10: Full hierarchy tree for task fMRI data. The subplots show the hemodynamic response for the cluster that converged at that stage.

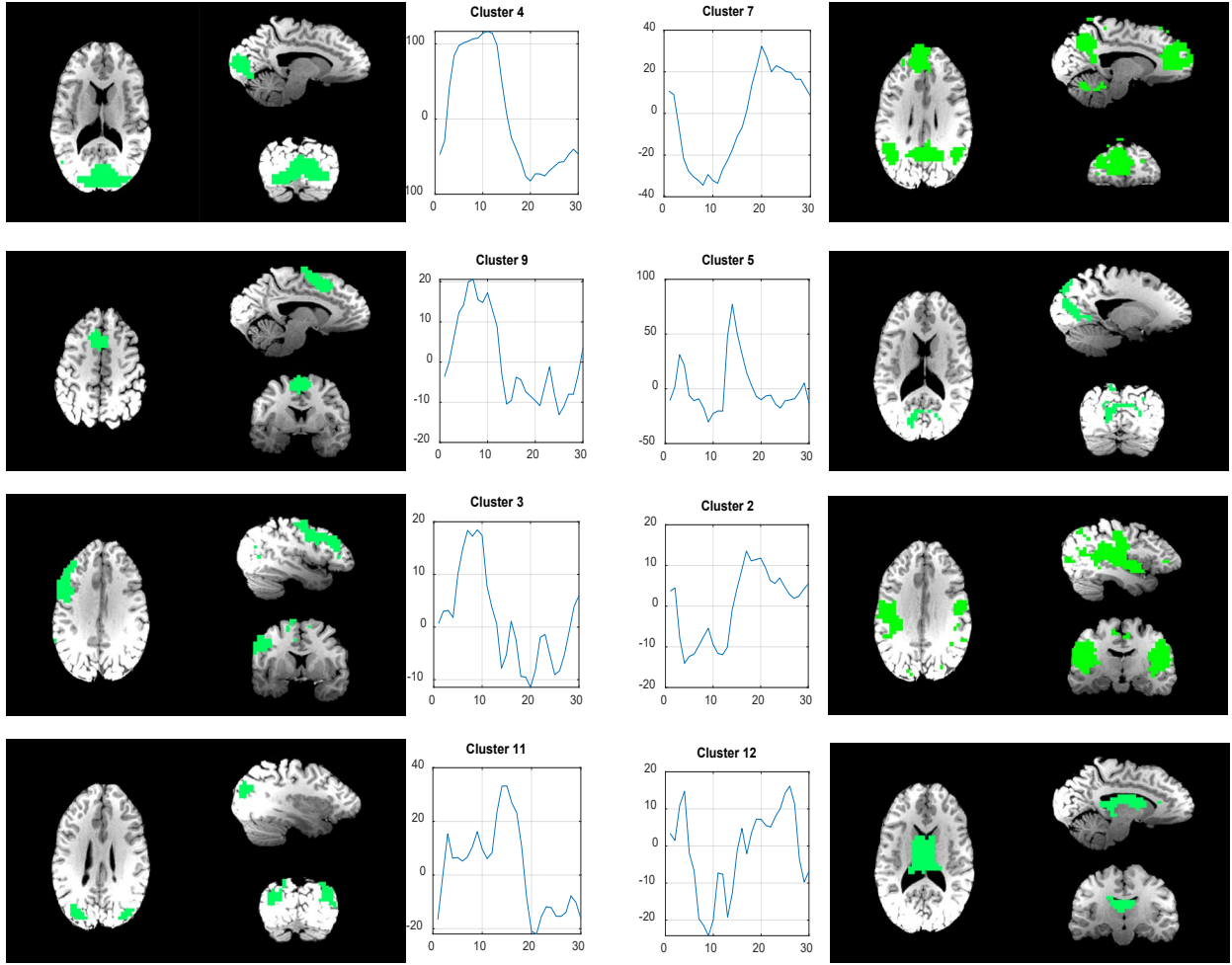

Figure 11: Spatial visualization of task fMRI clusters with corresponding hemodynamic response. The hemodynamic response for the brain regions involved in the task show a positive correlation with the experiment paradigm while the hemodynamic response in the default mode network, which is usually considered a task negative network, shows a negative correlation with the experiment paradigm.

## 4. Discussion

Spatial parcellation of brain volume based on functional similarity is of prime importance in fMRI data, especially for resting-state studies. Numerous techniques ranging from atlas-based to data-driven techniques, are used for this task. Each has its limitations and advantages. If using the atlas-based techniques, appropriate mapping and registration algorithms are of prime importance, because incorrect registration may yield inaccurate and mismatched spatial clusters. On the other hand, data-driven techniques do not necessarily require any such registration to standard atlas space but may yield spatial clusters that are spatially non-homogenous. Here, we have developed an unsupervised clustering technique that generates functionally homogenous spatial clusters. The functional homogeneity of the multistage clustering was also shown through the histogram of the mean intra-cluster values.

The hyperparameter in this approach gives an additional level of control over the size and homogeneity of the output clusters. Setting rigid hyperparameters yields a large number of small and functionally variant clusters while setting loose hyperparameters yields a few spatially large clusters. The effect of different hyperparameters has been briefly described in the earlier sections. In this section, we present an analysis of the effects of the two hyperparameters on the total number of output clusters. The number of stages (NS) in the hierarch broadly controls the total number of output clusters. However, at each node within the hierarchy, the k-means split might pass the convergence stopping criteria, and further splitting of the node is stopped. The relationship between the number of clusters and both the hyperparameters is shown in Table 1. The table shows the number of clusters obtained in a rest fMRI data set for different values of correlation threshold stopping criteria, from 0.3 to 0.95 in steps of 0.05, and the number of stages in the hierarchy, from 2 to 10. As expected, the number of clusters increases with an increase in correlation threshold, because fewer clusters converge early due to high correlation threshold. The number of clusters also increases with an increase in the number of stages in the hierarchy. As such, there is not an ideal set of hyperparameters that would work in all situation. Even with data driven techniques like ICA or atlas-based approach like the Craddock atlas, the number of parcellations depends on the specific application. That is the reason, Craddock atlas have different number of parcellations all the way from 10 to 1000. Moreover, even for ICA or simple k-means, the user needs to specify the desired number of brain ROIs. The hyperparameters in table 1 only gives an estimate of the hyperparameter range to get the desired number of functional ROIs.

Table 1: Number of clusters for different stages and correlation threshold

| CLUSTERS | | Stages in Hierarchy | | | | | | | | |
|---|---|---|---|---|---|---|---|---|---|---|
| | | 2 | 3 | 4 | 5 | 6 | 7 | 8 | 9 | 10 |
| Correlation Threshold | 0.3 | 3 | 4 | 5 | 6 | 7 | 7 | 7 | 7 | 8 |
| | 0.35 | 3 | 4 | 5 | 6 | 5 | 8 | 9 | 7 | 8 |
| | 0.4 | 3 | 4 | 5 | 7 | 8 | 10 | 9 | 10 | 11 |
| | 0.45 | 3 | 4 | 5 | 6 | 8 | 9 | 10 | 8 | 10 |
| | 0.5 | 3 | 4 | 5 | 7 | 8 | 9 | 10 | 10 | 9 |
| | 0.55 | 3 | 5 | 5 | 8 | 10 | 10 | 12 | 13 | 11 |
| | 0.6 | 3 | 5 | 7 | 11 | 13 | 16 | 22 | 22 | 19 |
| | 0.65 | 3 | 5 | 7 | 11 | 16 | 16 | 20 | 19 | 29 |
| | 0.7 | 3 | 5 | 7 | 11 | 17 | 23 | 30 | 32 | 43 |
| | 0.75 | 3 | 5 | 8 | 8 | 12 | 28 | 33 | 36 | 67 |
| | 0.8 | 4 | 7 | 10 | 16 | 24 | 45 | 47 | 74 | 100 |
| | 0.85 | 4 | 8 | 11 | 21 | 28 | 51 | 69 | 135 | 190 |
| | 0.9 | 4 | 8 | 13 | 19 | 35 | 54 | 87 | 170 | 234 |
| | 0.95 | 4 | 8 | 15 | 32 | 55 | 101 | 117 | 303 | 566 |

Apart from the hyperparameters, the effect of spatial smoothing was also studied for resting state data. Structural and functional similarity was computed on fMRI data preprocessed with a smoothing kernel with different FWHM. Figure 12 shows the comparison of spatial and functional similarity measures for different smoothing kernel size. The functional similarity was compared for ICA, multistage and simple k-means approach. The figure shows the plot of mean intra-cluster correlation value w.r.t FWHM. With increase in smoothing kernel size, the functional similarity increases for all the 3 clustering algorithms. Similar trend is observed in published ICA studies where a larger smoothing kernel increases functional coupling strength for ICA [Chen & Calhoun, 2018]. For large kernel, as average is taken over a larger volume, it becomes more similar. One more thing to observe is the reduction in the number of clusters with increasing FWHM. For higher values of FWHM, a smaller number of clusters is observed for the multistage approach. The structural similarity is also computed w.r.t FWHM. The figure shows the structural similarity, mSI value, between multistage clusters compared to different approaches, ICA, Craddock and anatomical atlas. Like functional similarity, the structural similarity also increases with increasing kernel size.

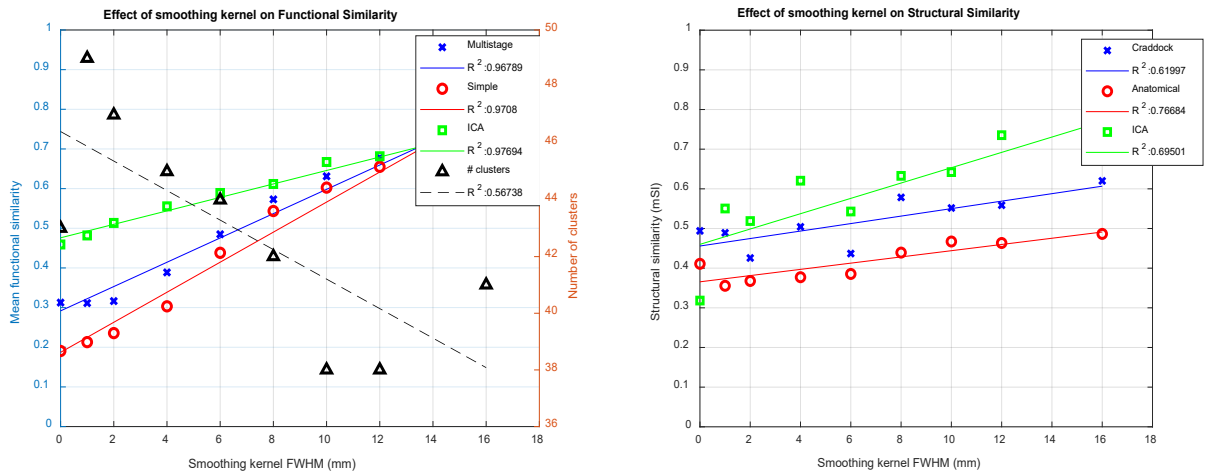

Figure 12: Comparison of (left) functional similarity, number of converged clusters and (right) structural similarity w.r.t smoothing kernel size. For functional similarity plot, different colors represent different clustering techniques while for functional similarity different colors indicate structural similarity of multistage approach with different clustering outputs. The figures show the absolute value for different structural and functional similarity measure and the straight-line fit indicating the trend. The $R^2$ goodness of fit score for the line is also indicated in the figure. The structural and functional similarity increases with increasing FWHM of the smoothing kernel while the total number of clusters that converge decreases.

The timing and memory requirements were analyzed analytically, and timing was also computed using simulation. The details of analytical analysis are given in the appendix A2. In summary, the hierarchical clustering performs faster only for smaller number of samples (<5000). The timing of hierarchical clustering increases exponentially with increasing number of samples while that of simple and multistage k-means increases linearly with increasing number of samples.

For task fMRI data, this clustering approach can also be used to explore the aspect of HR variability and whole-brain activation. The idea of whole-brain activation was introduced by Gonzales-Castillo et al. in 2012. In their paper, they showed that if the noise is suppressed by a significant level, which was achieved by averaging over 100 functional runs of the same experiment, activation is observed throughout the whole brain. The shape of the hemodynamic response, however, varies from region to region. Despite varying shape, all hemodynamic response shows a time-locked behavior to the experimental paradigm.

The results shown for the task fMRI analysis were from the same dataset. However, as opposed to averaging over 100 functional runs, the results shown were for a single randomly selected functional run. Figure 13 shows the comparison of different hemodynamic response obtained using the multistage k-means approach and the published results. Some of the hemodynamic response obtained by the clustering approach coincide highly with the published work. The hemodynamic responses which show time-locked correlation with the experimental paradigm are obtained by both approaches. The hemodynamic responses that do not match are the ones corresponding to noise and random temporal fluctuations.

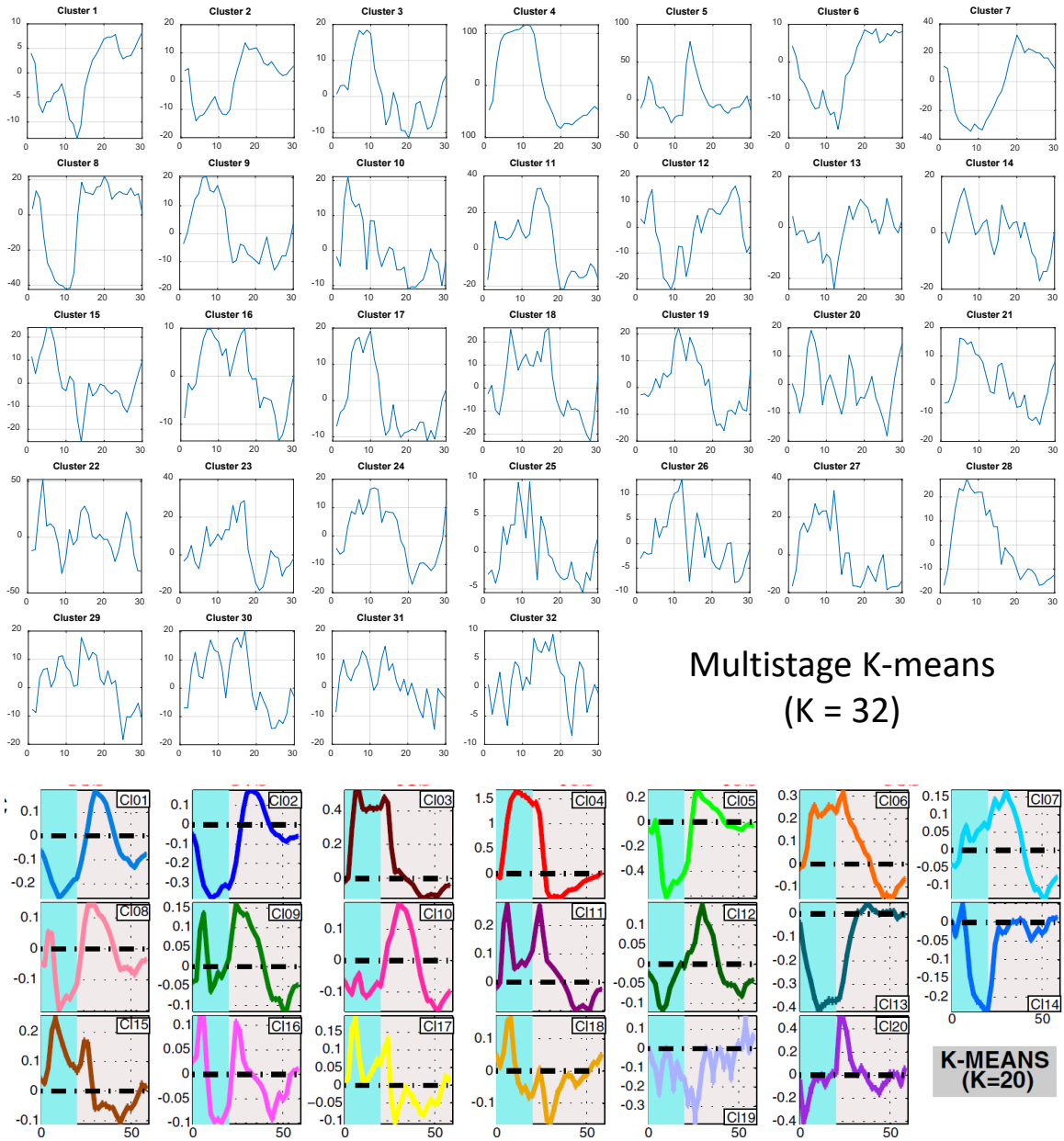

Figure 13: Hemodynamic response for different clusters/regions. [Top] Whole brain HR computed from average time series of the clusters obtained by multistage k-means algorithm. [Bottom] Whole brain HR published by Gonzales-Castillo et al. 2012. (Note: Scale for the HRs on the top is absolute intensity while for the HRs on the bottom is percent change)

## 5. Conclusions

The effectiveness of multistage k-means approach can be inferred from the results discussed above. Functional clusters can be obtained which show high spatial and functional homogeneity.

In the case of resting state fMRI, multistage approach can be used to obtain functional parcellation in an unsupervised manner without the use of any external atlas such as Craddock functional atlas or the Brodmann anatomical atlas. Just as with Craddock atlas, where we have functional parcels ranging from 10 to 1000, the proposed method can be used to obtain different numbers of parcellations depending on the choice of hyperparameters (correlation threshold and the number of stages). Currently, as it stands, the multistage approach is good for single subject functional parcellation and a direct point to point comparison between multiple subjects is still challenging. For group level analysis, functional mapping to standard atlas or other multi-subject functional parcellation schemes should be used [Thirion et al., 2006; Varoquaz et al., 2011]. In the case of task fMRI, it can be used to identify the primary and secondary activation regions. The average time series corresponding to different spatial clusters in task fMRI data are used to obtain hemodynamic response for different regions. Using this technique, we were also able to replicate the hemodynamic response corresponding to whole brain activation. As a future work, we wish to expand the utility of multistage clustering algorithm for group level studies. For task fMRI data, we plan to obtain hemodynamic response for different subjects, and investigate inter-subject hemodynamic response variability along with variability in the spatial region. Moreover, the multistage k-means technique itself is not limited to fMRI data. With some minor modifications this technique can be applied to any unsupervised clustering problem.

**Appendix A1: Dummy Example for Multistage k-means**

Figure 14 also shows the block diagram of multistage k-means clustering applied to a simple example. Here, the clustering approach is used with 3 stages of hierarchy, i.e. NS = 3. Consider that A is the collection of all feature vectors. In the first stage, k-means is used to split A into 2 clusters, B and C. Once that is done, the similarity is checked between B and C. The similarity is less than the stopping criteria, so we proceed to the next stage. In the second stage, again k-means is used to split both B and C into 2 clusters each. B is split into D and E while C is split into F and G. A high similarity is observed between D and E, shown in green, and hence it is likely that they are the subdivision of the same cluster, B in this case. Hence, B is considered as one independent cluster, shown in yellow. For F and G, a high similarity is not observed; thus, we move to the third stage. In the third stage F is split into H and I, and G is split into J and K. Again, a high similarity is observed between J and K, so G is considered as an independent cluster. Finally, as the maximum number of stages in the hierarchy is reached, H and I are considered as independent clusters. Thus, we end with 4 clusters from A, which are B, G, H and I.

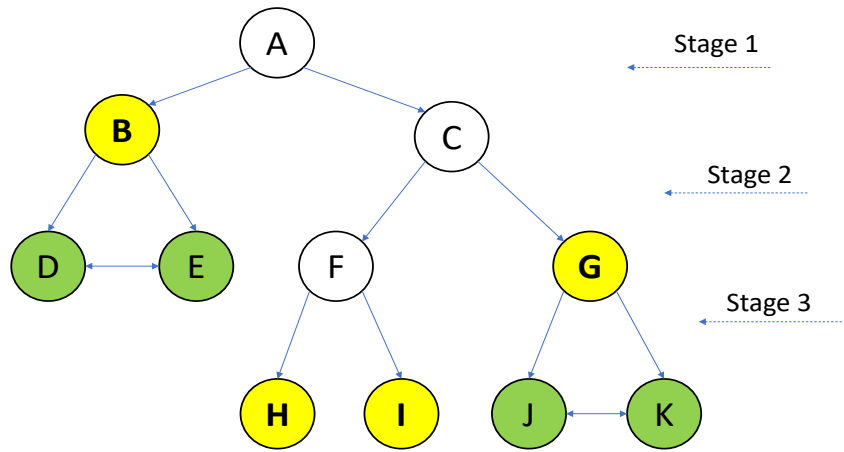

Figure 14: Dummy example for multistage k-means. Green circles show child clusters having similarity more than the threshold. Yellow clusters shows the converged clusters at the end of 3 stages.

**Appendix A2: Timing and Memory requirement comparison**

The timing and memory requirement performance of the multistage algorithm was analyzed on synthetic data and compared to both hierarchical and simple k-means clustering approach. First, the timing complexity is analyzed analytically. The operation that is repeated the greatest number of times during a clustering algorithm is the distance computation between pair of data points. During each iteration of simple k-means, distance is computed between data points and each cluster centroid. Thus, for 'N' data points and 'k' clusters, k·N distance computations are made for each iteration. Considering 'R' replicates with the total number of iterations per replicate as 'M', we get a total of (R·M·k·N) distance computations. For hierarchical clustering, a pairwise distance is computed just once for each pair of points. Thus, for N data points, we have an effective $\frac{(N(N-1))}{2}$ distance computations (considering symmetric N x N distance matrix with diagonal elements as zero).

For multistage algorithm, the worst-case scenario is when all the clusters split. For NS stages in the hierarchy, there can be a total of ($2^{NS}$-1) nodes. However, every node does not operate on all N data points. Let us consider that the parent node operates on N data points and splits the dataset into two clusters with 'n' and '(N-n)' samples. Thus, each child cluster only operates on a subset of the total data samples. Collectively, the total number of computations by all the children nodes is equivalent to the number of computations performed by the parent cluster provided other parameters are kept same. Therefore, the total computations in NS stages is equivalent to NS independent simple k-means computations with k = 2. Thus, the total number of distance computation is given by 2·NS·R·M·N.

The results obtained above suggests that the time complexity for simple and multistage k-means is O (n) while for hierarchical clustering is O ($n^2$). The worst case of multistage k-means will require less time than the hierarchical clustering for N>(4·NS·R·M) +1. The commonly used parameter values for multistage clustering, R = 5, M = 100, and NS = 7, yields N > 14,001 samples for it to perform better than hierarchical clustering. A typical fMRI data has anywhere from 50,000 to 200,000 voxels thus the above assumption holds true and the multistage approach requires lesser computation time than the hierarchical approach.

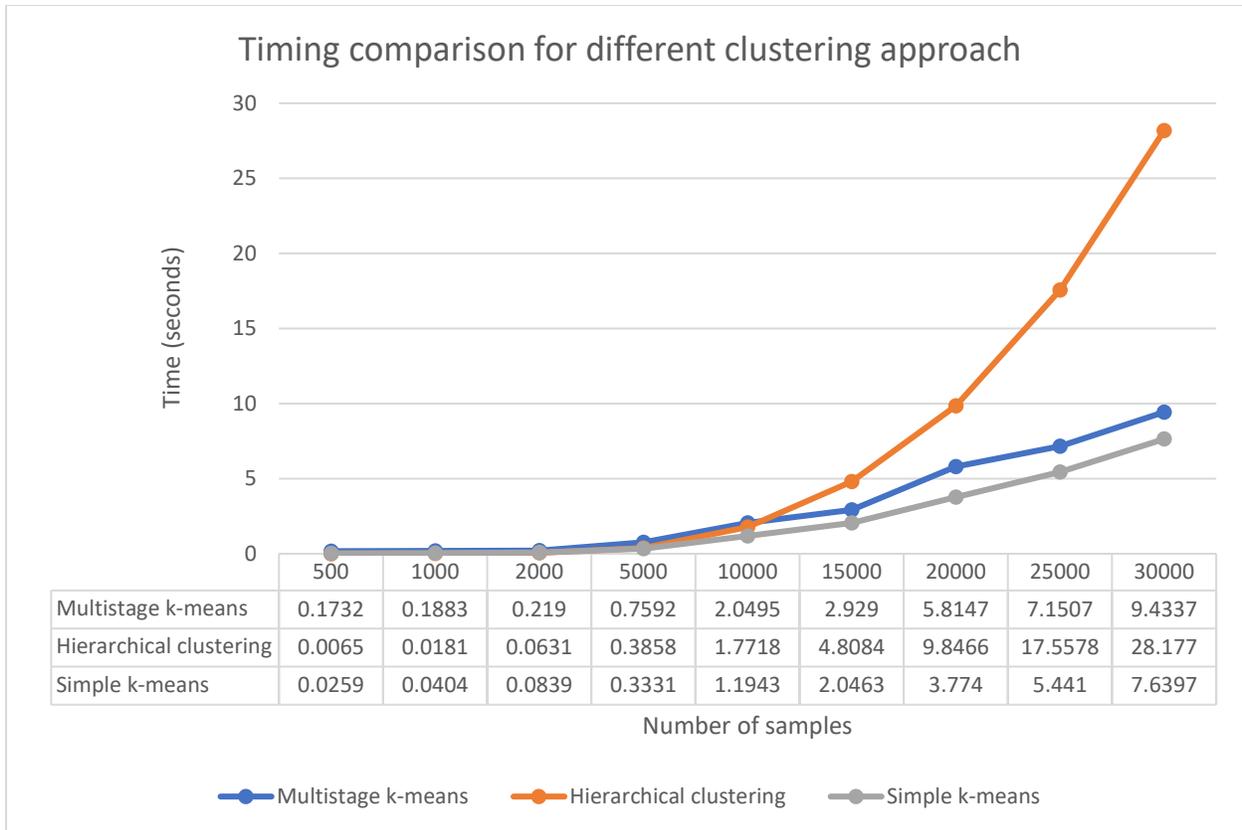

Figure 15: Timing performance for different clustering approach tested on simulated data. The hierarchical clustering follows a squared response where the time required is proportional to the square of the total number of samples. For simple and multistage k-means, the time required is a linear function of the total number of samples.

To have lower computation time than simple k-means, multistage approach needs to satisfy the condition: 2 NS<k. For the configuration mentioned above, for any k > 14, the multistage approach will perform faster than simple k-means which is reasonable assumption and most of the time the number of functional parcels is greater than twice the number of stages in hierarchy. Moreover, the space (memory) complexity for k-means is O (n) while for hierarchical clustering is O ($n^2$). Time performance was compared quantitatively for all three clustering methods using synthetic data. The synthetic data contains multiple random time series with 100 time points each. The time required to cluster the data was computed for different number of data samples (N). For each case, the algorithm was run 20 times and average time was calculated. Multiple runs reduce the chance of measuring outlier timing values. Figure 15 shows the plot of mean clustering time (in seconds) for multistage k-means, simple k-means, and hierarchical clustering for varying number of data samples. For lower number of samples, the hierarchical approach takes less time, but the time increases exponentially with increasing number of samples. Thus, the claim that the multistage approach performs better than hierarchical clustering for larger data samples is validated. The simple k-means does perform better with computation time than the multistage counterpart, but both the trends are similar and linear with increase in number of samples.

*Author Contributions:*

Conceptualization, Harshit Parmar; Formal analysis, Harshit Parmar; Funding acquisition, Sunanda Mitra; Methodology, Harshit Parmar; Project administration, Brian Nutter and Sunanda Mitra; Resources, Sunanda Mitra; Software, Harshit Parmar; Supervision, Brian Nutter, Rodney Long, Sameer Antani and Sunanda Mitra; Visualization, Harshit Parmar; Writing – original draft, Harshit Parmar; Writing – review & editing, Brian Nutter, Rodney Long, Sameer Antani and Sunanda Mitra.

*Funding:*

This research was supported in part by the Intramural Research Program of the National Institutes of Health (NIH), National Library of Medicine (NLM), and Lister Hill National Center for Biomedical Communications (LHNCBC), contract [HHSN276201800171P], and internal funding by Texas Tech University.

*Acknowledgments:*

We are grateful to Dr. Javier Gonzales-Castillo, staff scientist National Institute of Mental Health NIH, for providing the 100 functional run per subject dataset

[https://central.xnat.org/app/action/DisplayItemAction/search_value/100RunsPerSubj/search_element/xnat:projectData/search_field/xnat:projectData.ID]. We are also grateful to the Child Mind Institute and the International Neuroimaging Data-Sharing Initiative (IDNI) for providing the 1000 Functional Connectomes Project (FCP) dataset. [http://fcon_1000.projects.nitrc.org/index.html]

*Conflicts of Interest:*

The authors declare no conflict of interest.

**Harshit Parmar** received his BE degree in electronics engineering from Maharaja Sayajirao University of Baroda, India in 2016, and an MS degree in electrical engineering from Texas Tech University (TTU), Lubbock, Texas in 2018. He is currently pursuing his doctor of philosophy (PhD) in electrical engineering at TTU. He is working at the Biomedical Integrated Devices and Systems laboratory at TTU. His research interests include signal processing, pattern recognition, image processing, computer vision, and machine learning.

**Brian Nutter** received his BSEE and PhD degrees from Texas Tech University, Lubbock, Texas, in 1987 and 1990, respectively. He is an associate professor and associate chairman in the Department of Electrical and Computer Engineering, Texas Tech University. His interests include telecommunications, networks, biomedical signal and image processing, rapid prototyping, and real-time embedded systems.

**Rodney Long** leads a development group in creating applications for image-based biomedical information collection and dissemination. Currently, he is working in collaboration with the National Cancer Institute to develop a suite of tools for uterine cervix cancer databases. His research interests are in telecommunications, systems biology, image processing, and scientific/biomedical databases. He has an MA in applied mathematics from the University of Maryland.

**Sameer Antani** is currently (Acting) branch chief for the Communications Engineering Branch and the Computer Science Branch in the Lister Hill National Center for Biomedical Communications at the National Library of Medicine. He is a versatile lead researcher advancing the role of computational sciences and automated decision making in biomedical research, education, and clinical care. His research interests include topics in medical imaging and informatics, machine learning, data science, artificial intelligence, and global health. He applies his expertise in machine learning, biomedical image informatics, automatic medical image interpretation, data science, information retrieval, computer vision, and related topics in computer science and engineering technology.

**Sunanda Mitra** received her BSc degree in 1955 and MSc degree in 1957, both in physics, from Calcutta University, Calcutta, India, and her DSc (Doctors der Naturwissenshaften) degree in physics from Marburg University, Marburg, Germany, in 1966. She has served as the director of Computer Vision and Image Analysis Laboratory, Department of Electrical and Computer Engineering, Texas Tech University (TTU), Lubbock, Texas, since 1988. Prior to joining TTU in 1984, she worked as a research scientist at TTU's School of Medicine (1969 to 1983) and as a Visiting Faculty (1983 to 1984) at the Mount Sinai School of Medicine, New York. She also held a Faculty position (1960 to 1964) at Lady Brabourne College, Calcutta, India. She served on the Board of Scientific Counselors of the National Library of Medicine at the National Institutes of Health (USA) from 1997 to 2000. She holds the P.W. Horn Professorship at TTU and has published over 160 scientific articles including archival journal papers, invited papers, and book chapters. Her specialization includes medical image segmentation and analysis, data compression, 3-D modeling from stereo vision and pattern recognition. She has chaired the Technical Committee of Computational Medicine of the IEEE Computer Society.

**Caption List**

**Figure 1**: Pseudocode for the implementation of the multistage k-means algorithm.

**Figure 2**: (A) 2D ground truth matrix and visualization of synthetic 2D time series data. (B) Noisy time series with different SNR for voxels in different regions. (C), (D), (E) and (F) Results for synthetic time series data. Comparison of clustering output for multistage (right) and simple k-means (left) for different values of SNR, CT and NS. Different colors in the 2D matrix represents different clusters. The total number of clusters identified by the multistage algorithm is represented by 'K' on top of the plots. Mean similarity index values shown below each visualization.

**Figure 3**: (Left) Spatial clusters for multistage k-means. (Right) Craddock atlas wit 50 parcels. Different colors represent different clusters.

**Figure 4**: (left) structural similarity measure (mSI) between cluster outputs of different approach averaged (mean ± standard deviation) over 16 subjects. (right) t-value for comparing structural similarity for multistage k-means and different approach. The values in blue indicate significant higher values for multistage while in red indicate significant lower values for multistage approach as compared to others.

**Figure 5**: Spatial clustering comparison of resting-state fMRI data for (Top) simple k-means and (middle) multistage k-means and (bottom) ICA. Different colors represent different clusters.

**Figure 6**: Cluster overlap comparison for output of (left) multistage k-means, (center) simple k-means and (right) ICA. Clusters sorted according to gray matter overlap percentage. The x-axis represents normalized percentage with 0% - 100% scaled to 0 - 1.

**Figure 7**: Histogram for mean intra-cluster correlation for different clustering outputs. The average value for the histogram is mention on top of each subplot. The plots show the degree of functional similarity of different spatial clusters.

**Figure 8**: Extracted resting state brain networks using multistage k-means approach. The brain networks (red) are overlaid on the anatomical image (grayscale). The different brain networks shown here are (left to right, top to bottom) visual, default mode, primary motor, auditory, frontal, executive, sensorimotor and cerebellum.

**Figure 9**: Spatial clusters for task fMRI data obtained using multistage k-means. Different colors represent different clusters.

**Figure 10**: Full hierarchy tree for task fMRI data. The subplots show the hemodynamic response for the cluster that converged at that stage

**Figure 11**: Spatial visualization of task fMRI clusters with corresponding hemodynamic response. The hemodynamic response for the brain regions involved in the task show a positive correlation with the experiment paradigm while the hemodynamic response in the default mode network, which is usually considered a task negative network, shows a negative correlation with the experiment paradigm.

**Table 1**: Number of clusters for different stages and correlation threshold

**Figure 12**: Comparison of (left) functional similarity, number of converged clusters and (right) structural similarity w.r.t smoothing kernel size. For functional similarity plot, different colors represent different clustering techniques while for functional similarity different colors indicate structural similarity of multistage approach with different clustering outputs. The figures show the absolute value for different structural and functional similarity measure and the straight-line fit indicating the trend. The R2 goodness of fit score for the line is also indicated in the figure. The structural and functional similarity increases with increasing FWHM of the smoothing kernel while the total number of clusters that converge decreases.

**Figure 13**: Hemodynamic response for different clusters/regions. [Top] Whole brain HR computed from average time series of the clusters obtained by multistage k-means algorithm. [Bottom] Whole brain HR published by Gonzales-Castillo et al. 2012. (Note: Scale for the HRs on the left is absolute intensity while for the HRs on the right is percent change)

**Figure 14:** Dummy example for multistage k-means. Green circles show child clusters having similarity more than the threshold. Yellow clusters shows the converged clusters at the end of 3 stages.

**Figure 15:** Timing performance for different clustering approach tested on simulated data. The hierarchical clustering follows a squared response where the time required is proportional to the square of the total number of samples. For simple and multistage k-means, the time required is a linear function of the total number of samples.